\documentclass[sigplan,10pt,screen]{acmart}
\renewcommand\footnotetextcopyrightpermission[1]{}
\AtBeginDocument{%
  }

\setcopyright{acmlicensed}
\copyrightyear{2018}
\acmYear{2018}
\acmDOI{XXXXXXX.XXXXXXX}
\acmConference[Conference acronym 'XX]{Make sure to enter the correct
  conference title from your rights confirmation email}{June 03--05,
  2018}{Woodstock, NY}
\acmISBN{978-1-4503-XXXX-X/2018/06}

\usepackage[most]{tcolorbox}

\usepackage{multicol}
\usepackage{multirow}
\usepackage{graphicx}
\usepackage{makecell}
\usepackage{enumitem}
\usepackage{cleveref}

\newcommand{\Reb}[1]{\textcolor{red}}

\usepackage{booktabs}
\usepackage{xspace}
\usepackage[T1]{fontenc}

\newcommand{\proteus}{\emph{Optimus}\xspace}

\acmSubmissionID{123-A56-BU3}

\begin{document}

\title{Optimus: Elastic Decoding for Efficient \\  Diffusion LLM Serving}

\author{Chiyue Wei}
\authornote{Both authors contributed equally to this research. \\ Corresponding author: Cong Guo $<$cong.guo@duke.edu$>$}
\email{chiyue.wei@duke.edu}
\affiliation{%
  \institution{Duke University}
  \city{Durham}
  \state{North Carolina}
  \country{USA}
}

\author{Cong Guo}
\authornotemark[1]
\email{cong.guo@duke.edu}
\affiliation{%
  \institution{Duke University}
  \city{Durham}
  \state{North Carolina}
  \country{USA}
}

\author{Bowen Duan}
\email{bowen.duan@duke.edu}
\affiliation{%
  \institution{Duke University}
  \city{Durham}
  \state{North Carolina}
  \country{USA}
}

\author{Junyao Zhang}
\email{junyao.zhang@duke.edu}
\affiliation{%
  \institution{Duke University}
  \city{Durham}
  \state{North Carolina}
  \country{USA}
}

\author{Haoxuan Shan}
\email{haoxuan.shan@duke.edu}
\affiliation{%
  \institution{Duke University}
  \city{Durham}
  \state{North Carolina}
  \country{USA}
}

\author{Yifei Wang}
\email{yifei.wang4@duke.edu}
\affiliation{%
  \institution{Duke University}
  \city{Durham}
  \state{North Carolina}
  \country{USA}
}

\author{Yangjie Zhou}
\email{yj_zhou@nus.edu.sg}
\affiliation{%
  \institution{National University of Singapore}
  \city{Singapore}
  \country{Singapore}
}

\author{Hai ``Helen'' Li}
\email{hai.li@duke.edu}
\affiliation{%
  \institution{Duke University}
  \city{Durham}
  \state{North Carolina}
  \country{USA}
}

\author{Danyang Zhuo}
\email{danyang@cs.duke.edu}
\affiliation{%
  \institution{Duke University}
  \city{Durham}
  \state{North Carolina}
  \country{USA}
}

\author{Yiran Chen}
\email{yiran.chen@duke.edu}
\affiliation{%
  \institution{Duke University}
  \city{Durham}
  \state{North Carolina}
  \country{USA}
}



\begin{abstract}
Large language model (LLM) serving is fundamentally limited by inefficient hardware utilization. Autoregressive (AR) decoding underutilizes GPUs due to its strictly sequential execution, while diffusion LLMs (DLLMs) improve throughput by decoding multiple tokens per iteration. However, fixed block-size diffusion decoding exhibits strong load sensitivity: large blocks exploit idle GPU resources under low load, but saturate early and incur substantial redundant computation under high load. As a result, throughput gains vanish beyond saturation, and no single decoding granularity performs well across dynamic serving workloads.

We present \textbf{\proteus}, a serving system that enables elastic decoding for diffusion LLMs by dynamically adapting decoding granularity to runtime load. The key idea is to treat decoding granularity as a runtime control variable, balancing GPU utilization and token efficiency. \proteus combines chunked decoding, which enables fine-grained execution without retraining, with saturation-aware scheduling, a closed-loop mechanism that selects chunk sizes based on runtime conditions. Together with system-level optimizations and customized attention kernels, \proteus achieves significant performance improvements while preserving model accuracy. Experiments show that \proteus delivers up to $6.1\times$ throughput improvement over AR decoding and $4.3\times$ improvement over fixed-block diffusion LLM, while maintaining stable performance across diverse load regimes and improving end-to-end serving capacity under latency constraints. The source code is available at \url{https://github.com/dubcyfor3/Optimus}.
\end{abstract}

\settopmatter{printfolios=true}
\maketitle
\pagestyle{plain}

\section{Introduction}
Large language models (LLMs)~\cite{devlin2019bert,dettmers2022gpt3,achiam2023gpt,yang2025qwen3,jiang2023mistral7b} have rapidly become critical infrastructure for a wide range of applications, including conversational~\cite{achiam2023gpt,team2023gemini,guan2026evaluating}, code generation~\cite{danyaro2025llm,chen2021evaluating}, and data analysis~\cite{nasseri2023applications}. As LLM-based services scale, serving efficiency has emerged as a primary systems bottleneck. Despite the massive computational capability of modern GPUs, autoregressive (AR)~\cite{vaswani2017attention,zeng2023transformers} decoding often fails to utilize hardware resources effectively.

In AR decoding, tokens are generated sequentially, one token per iteration. When the batch size (bs) is small (e.g., $\text{bs}=1$), execution degenerates into a sequence of GEMV (matrix-vector multiplication) operations with poor weight reuse, making it inherently memory-bound and leaving most GPU compute units idle. 
On an A100 GPU running Qwen-8B~\cite{yang2025qwen3}, we observe utilization dropping below 1\% under small-batch workloads. 
Continuous batching (CB)~\cite{yu2022orca,holmes2024deepspeed,kwon2023vllm, zheng2024sglang}  has been widely adopted to mitigate this inefficiency by dynamically batching requests to increase the effective GEMM (matrix multiplication) dimension, improving arithmetic intensity and boosting throughput without significantly increasing latency. As a result, CB has become a de facto standard in modern LLM serving systems.

\begin{figure}[t]
    \centering
    \includegraphics[width=\linewidth]{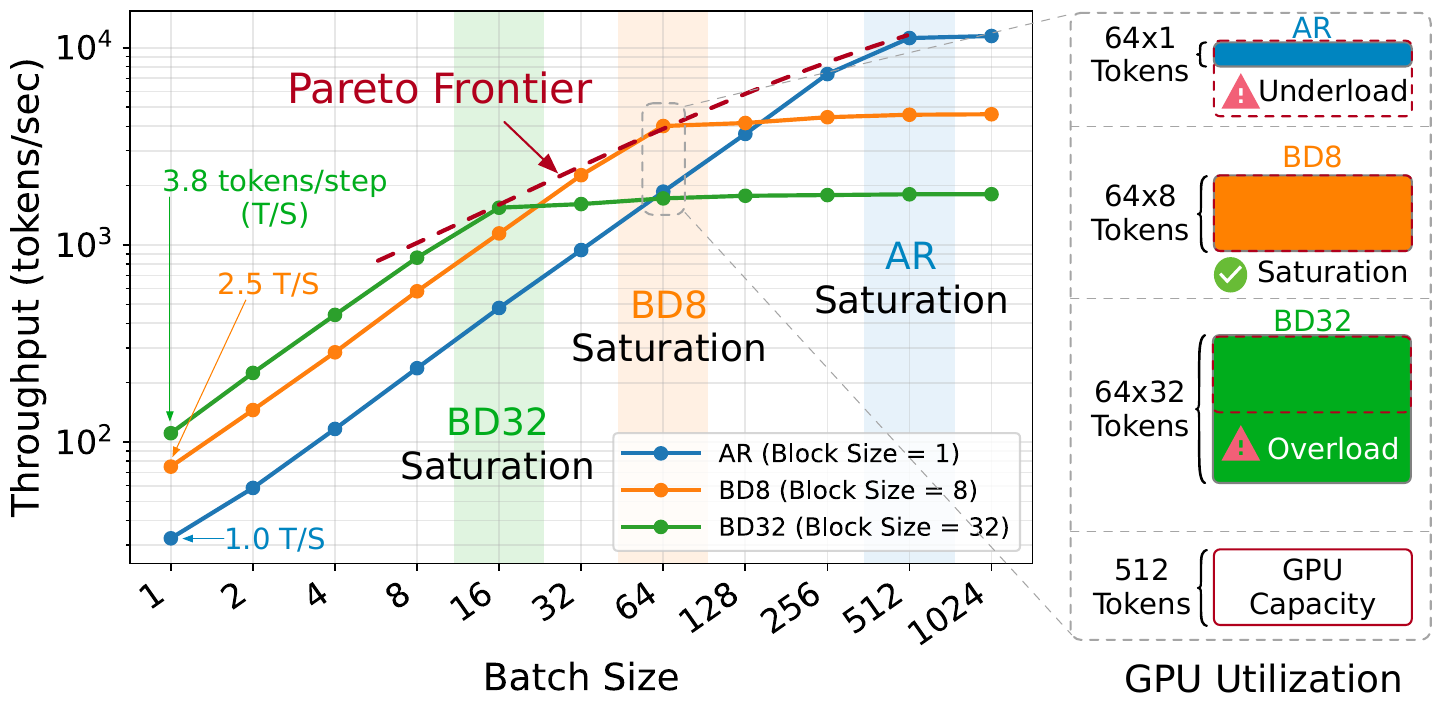}
    
\vspace{-4mm}
\caption{
\textbf{Load sensitivity of fixed-granularity decoding.}
Throughput under increasing load for autoregressive (AR) (Qwen-8B~\cite{yang2025qwen3}) and block diffusion (BD) (SDAR-8B~\cite{cheng2025sdar} with block sizes 8 and 32) on an A100 80GB GPU.
}
\vspace{-6mm}
    \label{fig:intro}
    \Description{}
\end{figure}

However, CB only exploits \emph{inter-request parallelism}. When system load is low, GPU underutilization persists due to the intrinsic token-by-token execution granularity of AR decoding, which fundamentally limits the amount of parallel work exposed per request. To characterize this behavior, we benchmark throughput across a range of batch sizes. As shown in \Cref{fig:intro}, AR decoding remains underloaded at small batch sizes and only approaches saturation under unrealistically high concurrency (e.g., $\text{bs}=512$).

Recent advances in diffusion-style large language models (DLLMs)~\cite{nie2025large,zhu2025llada,bie2025llada2,arriola2025block,cheng2025sdar,li2022diffusion} increase decoding granularity by computing multiple tokens per iteration. By operating on blocks of tokens instead of single-token steps, diffusion decoding exposes significantly more {\emph{intra-request parallelism}} and can effectively utilize otherwise idle GPU resources.
As shown in \Cref{fig:intro}, under low-load conditions, larger block sizes achieve substantially higher throughput (3.8 tokens/step) than AR decoding (1.0 token/step). By evaluating multiple token positions simultaneously, the model can commit high-confidence tokens earlier, accelerating decoding. However, this improvement comes with two key limitations.

First, diffusion decoding tends to saturate the GPU at relatively small batch sizes. As shown in \Cref{fig:intro}, a configuration with block size 32 (BD32) reaches saturation around $\text{bs}=16$. Beyond this point, increasing the load no longer improves throughput, as the GPU is already fully utilized.
Second, diffusion decoding exhibits low token utilization. For example, with a block size of 32, the model computes 32 tokens per step but only commits around 3.8 tokens on average. In contrast, AR decoding achieves full utilization, where each computed token is committed. This gap indicates that a large fraction of computation in diffusion decoding is redundant.
While redundant computation is acceptable under underloaded conditions, it becomes pure overhead under GPU saturation. Consequently, large-block diffusion decoding loses its advantage and may underperform smaller block sizes  (e.g., BD8 at $\text{bs}>32$) or even AR decoding at higher load (e.g., $\text{bs}>128$).

This behavior reveals a fundamental limitation of fixed block-size decoding: a single configuration cannot perform well across all load regimes. In dynamic serving environments, fixed-granularity decoding inevitably leads to suboptimal performance.
To address this limitation, decoding granularity must adapt to runtime load. By dynamically adjusting block size, the system can operate near the optimal point along the \textbf{\emph{Pareto frontier}} between GPU utilization and token utilization, achieving consistently high performance.

We present \textbf{\proteus}, a serving system that enables elastic decoding guided by GPU saturation for diffusion large language models. \proteus dynamically adjusts decoding granularity at runtime to keep GPU utilization near saturation while controlling token decoding efficiency.

The core insight behind \proteus is an \emph{algorithm--system trade-off}. Algorithmically, token utilization varies with decoding granularity (e.g., BD32 commits 3.8 tokens per 32 computed tokens, while AR approaches 100\%). System-wise, larger blocks expose greater parallelism and improve GPU utilization. By jointly considering these factors, \proteus enables dynamic granularity selection that achieves substantially higher efficiency than fixed configurations. Realizing this idea introduces two key challenges:

\textbf{(1) Enabling Chunked Granularity.}
Existing diffusion models operate at fixed block sizes (e.g., BD8, BD32), which are trained as separate models with different parameters. In practice, serving systems cannot switch between multiple models at runtime. To address this, \proteus introduces \emph{chunked decoding}, a system-level mechanism that decomposes a diffusion decoding block into fine-grained execution units without retraining multiple models.

\textbf{(2) Runtime Elastic Scheduling.}
Even with flexible granularity, selecting the appropriate chunk size at runtime remains challenging. The system must continuously sense GPU utilization and token efficiency under dynamic workloads, while minimizing the overhead of switching execution granularity. To address this, \proteus implements \emph{elastic scheduling}, a closed-loop control mechanism that dynamically selects chunk sizes based on runtime conditions, maintaining execution near the GPU saturation point and avoiding both underutilization and overload.

\begin{figure*}[t]
    \centering
    \includegraphics[width=0.99\linewidth]{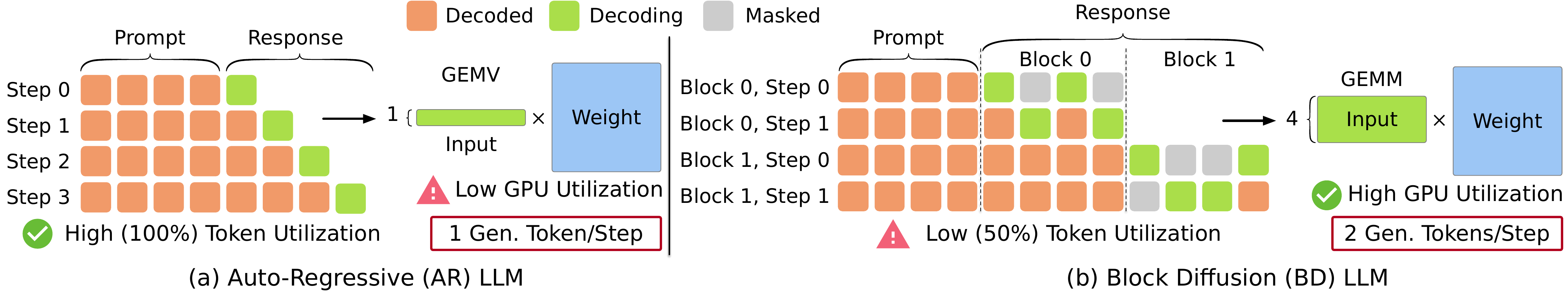}
    \vspace{-5mm}
    \caption{
    Comparison between autoregressive decoding (left) and block diffusion decoding (right).
    AR generates one token per step, while diffusion-style decoding generates multiple tokens per iteration, batch size 1.
    }
    \vspace{-3mm}
    \Description{}
    \label{fig:ar_vs_dllm}
\end{figure*}

\proteus integrates chunked decoding with a saturation-aware scheduling framework, together with system-level optimizations and customized attention kernels. This design enables efficient execution without sacrificing model accuracy.
Our evaluation shows that, under the same system configuration, \proteus achieves significant throughput gains over both AR and fixed-block diffusion decoding. On 8B-scale models, \proteus improves throughput by an average of $2.1\times$ (up to $6.1\times$) over AR, and $1.3\times$ (up to $4.3\times$) over BD32. Under realistic serving workloads with a service-level objective (SLO) constraint, \proteus further improves end-to-end serving capacity by up to $3.5\times$ over AR and up to $2.0\times$ over BD32, and $11.8\times$ over SGLang DLLM implementations.

These results demonstrate that \proteus makes diffusion LLM serving practical under dynamic workloads and provides a system-level foundation that may inform future algorithm design.
This study makes the following contributions:

\begin{itemize}[noitemsep, topsep=2pt, leftmargin=*]
    \item We identify and characterize the \emph{load sensitivity} problem in diffusion LLM serving, and formalize the trade-off between token utilization and GPU utilization.
    \item We propose \emph{chunked decoding}, a system-level mechanism that enables fine-grained, runtime-adjustable diffusion decoding without retraining.
    \item We develop \emph{saturation-aware elastic scheduling}, a closed-loop control framework that dynamically adapts decoding granularity to runtime GPU load.
    \item Through extensive evaluation, we demonstrate throughput improvement and serving capacity improvement over AR decoding over fixed-block diffusion decoding.
\end{itemize}

\section{Background}


\subsection{LLM and Autoregressive Decoding}
Most LLMs adopt the transformer architecture~\cite{vaswani2017attention} and generate text autoregressively (AR)~\cite{yang2025qwen3,achiam2023gpt,dettmers2022gpt3}, which predicts each token conditioned on the prompt and previously generated tokens. As illustrated in \Cref{fig:ar_vs_dllm} (left), the prompt is fully available at the start (step 0), whereas response tokens are produced strictly one by one. At decoding step $t$, the model takes the prompt and generated prefix (orange) as input and produces a single token $x_t$ (green), which is then appended to the prefix for the next step.

This token-by-token dependency defines AR decoding and fundamentally limits its hardware efficiency. As a result, a request that generates $T$ output tokens typically requires $T$ sequential decoding iterations, which makes latency grow linearly with output length. For interactive workloads such as chat and code completion, where batch size is often small, and latency is critical, this sequential execution becomes the dominant systems bottleneck~\cite{shazeer2019fast, yu2022orca}

\subsection{GPU Underutilization and Serving Optimizations}
From a system perspective, AR decoding is notoriously inefficient on modern GPUs. At small batch sizes, each decoding step effectively resembles a set of narrow matrix--vector (GEMV) or ``skinny'' matrix multiplications, rather than compute-dense matrix--matrix multiplications (GEMM). These kernels exhibit low arithmetic intensity and limited weight reuse, making performance memory-bandwidth-bound rather than compute-bound~\cite{shazeer2019fast, nvidia-gemm, roofline}. 
This mismatch explains why single-request decoding frequently fails to fully utilize GPU compute resources.

A standard systems-level remedy is \emph{continuous batching} (CB), also known as iteration-level scheduling~\cite{yu2022orca, kwon2023vllm, zheng2024sglang}. Instead of executing one request at a time, the server dynamically merges decoding steps from multiple concurrent requests into a shared batch. Because different requests may be at different decoding positions, the scheduler continuously admits new sequences and retires finished ones, thereby maintaining a larger effective batch over time. This enlarged batch turns many decoding kernels from GEMV-like execution into larger GEMM-like execution, improving arithmetic intensity and increasing hardware occupancy.

Continuous batching has become a key design principle in modern LLM serving engines because it improves throughput substantially without incurring the large head-of-line blocking typical of static batching~\cite{yu2022orca, kwon2023vllm}. Nevertheless, CB exploits only \emph{``inter-request''} parallelism. Its effectiveness depends on workload concurrency. Under low-load conditions, when there are too few simultaneous requests to form a large batch, AR decoding remains fundamentally sequential within each request. Consequently, GPU underutilization is alleviated but not eliminated: efficiency improves only when sufficient request-level concurrency is available.

\begin{figure*}[t]
    \centering
    \includegraphics[width=\linewidth]{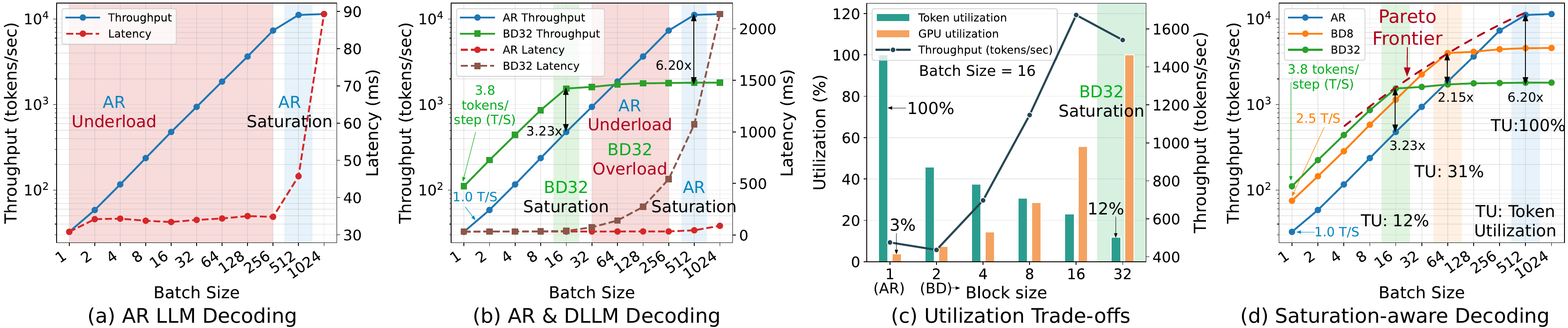}
    \vspace{-8mm}
\caption{
Motivation for saturation-aware decoding. 
(a) AR underutilizes the GPU. 
(b) Diffusion improves utilization but suffers under high load. 
(c) Granularity trades off GPU and token utilization. 
(d) The optimal point tracks the saturation boundary.
Experiments use Qwen3-8B (AR) and SDAR-8B (multiple block sizes, from 2 to 32) on an A100 80GB.
}
    \vspace{-2mm}
    \label{fig:motivation}
    \Description{}
\end{figure*}

\subsection{Diffusion LLM and Block Diffusion}

To fundamentally overcome the sequential bottleneck of AR decoding, diffusion-style large language models (DLLMs)~\cite{zhu2025llada,bie2025llada2,arriola2025block,cheng2025sdar} offer a superior alternative. As illustrated on the right side of \Cref{fig:ar_vs_dllm}, block diffusion decoding abandons the token-by-token progression of AR models. Instead, it predicts and refines a block of tokens jointly, allowing the model to make progress across multiple positions simultaneously~\cite{arriola2025block}. 
During decoding, tokens whose confidence exceeds a threshold are committed in each iteration.

This blockwise execution represents a paradigm shift in generation efficiency. By generating multiple tokens per iteration, Diffusion LLMs drastically increase \emph{``intra-request''} parallelism. Each decoding round performs substantially more useful work, transforming narrow operations into wider, more compute-dense executions that inherently align with the throughput-oriented architecture of modern GPUs. Consequently, Diffusion LLMs maintain high hardware utilization and deliver significantly faster generation speeds even at a batch size of 1, effectively bypassing the concurrency reliance that limits AR models.

While diffusion-style decoding offers clear system benefits, realizing robust end-to-end efficiency requires careful management. Block-based decoding introduces a granularity tradeoff: larger blocks increase parallelism and peak throughput, but reduce runtime flexibility and make execution more sensitive to load variations. Fixed-block diffusion approaches may excel under specific conditions while degrading in others. This motivates system support for dynamically controlling decoding granularity, which is the focus of this work.

\section{Motivation}

We begin with autoregressive (AR) decoding to illustrate GPU underutilization, and then compare it with diffusion decoding (BD32) to show how larger decoding granularity improves utilization while introducing new limitations. These observations reveal a trade-off between GPU utilization and token utilization, motivating a system design that dynamically adapts decoding granularity. We conclude by outlining the key challenges in realizing such a design.

\subsection{AR Decoding Fails to Saturate the GPU}

\Cref{fig:motivation}(a) shows throughput and latency as the batch size increases. Under small batch sizes, AR decoding severely underutilizes the GPU: each decoding step exposes only limited parallel work, leaving substantial hardware capacity idle. In this underloaded regime, increasing the batch size improves throughput almost proportionally, while latency remains nearly unchanged. This suggests that additional requests can be absorbed using otherwise idle GPU resources, yielding what is effectively a near-free throughput gain.
This observation is well known in prior LLM serving systems and motivates continuous batching, which improves hardware efficiency by increasing the effective batch size~\cite{yu2022orca,kwon2023vllm,zheng2024sglang}.

More importantly, because AR decoding has a fixed granularity of one token per step, it is intrinsically difficult for a single request to expose enough parallel work to saturate the GPU. In our example, AR decoding on an A100 does not approach saturation until the batch size reaches around 512. Such concurrency is rarely sustainable in real-world serving, especially under low-load conditions. As a result, even with continuous batching, AR decoding often leaves substantial GPU resources idle in practice.

\textbf{Takeaway.}
AR decoding rarely saturates the GPU. With one-token-per-step granularity, it requires unrealistically high concurrency to fully utilize modern accelerators, leaving substantial capacity idle in practical serving workloads.

\subsection{DLLM Improves Utilization but Saturates Early}
\Cref{fig:motivation}(b) compares AR decoding with block diffusion using a block size of 32 (BD32). By performing more computation per step, BD32 converts otherwise idle GPU resources into useful decoding progress, achieving an average of 3.8 tokens per step. As a result, under small batch sizes (e.g., $bs \leq 16$), BD32 achieves up to $3.2\times$ higher throughput than AR, while maintaining comparable latency. This demonstrates that DLLMs can effectively improve throughput with little to no additional latency cost in the underloaded regime.

However, increasing decoding granularity also increases the total computation per step. BD32 reaches GPU saturation at relatively small batch sizes, after which additional workload leads to queuing and rapidly increasing latency. In this regime, redundant computation dominates execution, and throughput no longer improves. In contrast, AR decoding continues to scale at higher batch sizes and eventually outperforms BD32, with up to $6.2\times$ higher throughput.

\textbf{Takeaway.}
Diffusion decoding improves GPU utilization under low load but saturates earlier. As a result, its performance advantage is limited to a narrow operating region.

\subsection{Algorithm-System Trade-off}
The observations above suggest that decoding granularity controls the GPU saturation point. By adjusting the block size, diffusion decoding trades off GPU utilization against token-level efficiency. To formalize this trade-off, we distinguish two metrics: \emph{GPU utilization} (GU), which reflects hardware usage, and \emph{token utilization} (TU), which reflects the fraction of useful computation.

GPU utilization depends on the amount of parallel work exposed to the accelerator. We define the \emph{effective workload} (EW) per decoding step as:
$
\text{EW} = \text{batch size} \times \text{block size}.
$
Increasing either term pushes execution toward the GPU saturation point. In practice, for an A100 GPU, saturation is reached when EW approaches a hardware-dependent limit (e.g., around 512 in our setup). In contrast, TU captures algorithmic efficiency:
\vspace{-5pt}
\begin{equation*}
    \text{TU} = \frac{\text{\# committed tokens}}{\text{\# computed tokens}}
\end{equation*}
AR decoding achieves $\text{TU}=100\%$, while diffusion decoding introduces redundant computation as block size increases (e.g., BD32 commits only $3.8/32 \approx 12\%$ token per step).

These objectives are inherently in tension: larger block sizes improve GPU utilization but reduce TU. As shown in \Cref{fig:motivation}(c), no single block size maximizes both, and different block sizes correspond to different trade-off points.

\textbf{Takeaway.}
Neither AR nor large-block diffusion alone achieves optimal efficiency, making dynamic control over decoding granularity necessary.

\subsection{Saturation-Aware Frontier}
The performance of diffusion decoding depends jointly on batch size and block size. As shown in \Cref{fig:motivation}(d), different block sizes achieve optimal performance under different load conditions.
Under low load, large blocks improve throughput by exploiting idle GPU capacity. Under high load, however, the GPU approaches saturation, and redundant computation from large blocks degrades efficiency. As a result, each block size has a distinct operating region, and no single fixed granularity performs well across all regimes.

This behavior gives rise to a \emph{saturation-aware frontier}: the optimal decoding strategy lies near the GPU saturation boundary, and the best block size shifts with load. Therefore, achieving consistently high performance requires dynamically adapting decoding granularity to follow this frontier.

\textbf{Takeaway.}
The optimal decoding granularity is not fixed; it varies with runtime load. Static block sizes are inherently suboptimal in dynamic serving environments.

\subsection{Challenges}

Realizing saturation-aware decoding requires dynamically tracking the saturation frontier at runtime. However, this is fundamentally challenging, as it requires both flexible decoding granularity and load-aware scheduling, neither of which is supported by existing designs.

\textbf{(1) Enabling Chunked Decoding.}  
A straightforward approach to supporting multiple block sizes is to train separate models (e.g., BD2, BD4, BD8, BD16, BD32 of SDAR-8B~\cite{cheng2025sdar}), as illustrated in \Cref{fig:motivation}(c)(d) and adopted by existing diffusion LLM approaches. Each model operates at a fixed decoding granularity with independently trained parameters. However, this design is impractical at serving time due to prohibitive memory and model-switching overhead.

Instead, we must enable multiple granularities \emph{within a single model} by decoupling execution granularity from the model's original block size. We adopt a large-block model as the base and introduce a finer-grained abstraction, \textbf{\emph{chunks}}, to represent sub-block execution units. This leads to  \textbf{\emph{chunked decoding}}, which enables flexible control over decoding granularity without requiring multiple models.

However, chunked decoding is non-trivial. Unlike chunked prefill~\cite{agrawal2023chuckprefill,zhong2024distserve}, DLLM decoding involves strict intra-block dependencies, including token ordering constraints and KV-cache dependencies. Achieving both correctness and efficiency under such fine-grained execution requires rethinking the decoding process. \Cref{sec:decoding} presents our solution.

\textbf{(2) Scheduling for Elastic Decoding.}  
Even with chunked decoding, efficiently selecting the optimal chunk size at runtime remains challenging, as it depends on system load and must adapt dynamically to track the saturation frontier.

This requires the system to (i) sense GPU utilization in real time, (ii) select chunk sizes that balance GPU utilization and token utilization, and (iii) transition between granularities without introducing instability or overhead. Unlike static batching or fixed-block decoding, this forms a closed-loop scheduling problem under dynamic and unpredictable workloads. Designing an \textbf{\emph{elastic decoding scheduler}} that is both responsive and stable is essential for achieving consistently high performance. \Cref{sec:scheduling} presents our design.

\section{Streaming Chunked Decoding}\label{sec:decoding}

\begin{figure*}[t]
    \centering
    \includegraphics[width=\linewidth]{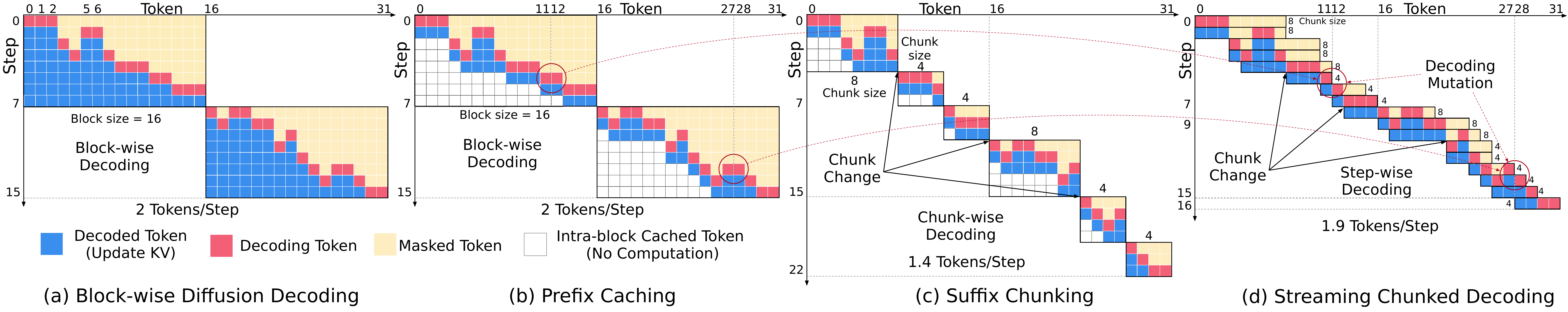}
    \vspace{-8mm}
\caption{
Chunked decoding overview. 
(a) Block-wise diffusion decoding. 
(b) Prefix caching removes prefix dependency. 
(c) Suffix chunking enables fine-grained execution. 
(d) Streaming chunked decoding restores execution order and efficiency.
}
    \vspace{-3mm}
    \Description{}
    \label{fig:chunk}
\end{figure*}

This section presents \emph{chunked decoding}, a mechanism that enables fine-grained execution within diffusion LLMs while preserving correctness. Starting from block-wise decoding, we progressively remove intra-block dependencies and introduce chunk-level control to enable flexible execution.

\subsection{Original Block Diffusion}

We first describe the execution model of block-wise diffusion (BD) decoding, shown in \Cref{fig:chunk}(a).
\begin{table}[b]
\centering
\vspace{-4mm}
\caption{Token states in block-wise diffusion decoding.}
\vspace{-4mm}
\label{tab:states}
\small
\resizebox{1\linewidth}{!}
{\renewcommand{\arraystretch}{1.2}    
\begin{tabular}{l |l l l}
\toprule
 & \textbf{Masked} & \textbf{Decoding} & \textbf{Decoded} \\
 & (yellow) & (red) & (blue) \\
\midrule
\textbf{Input} 
& mask token 
& mask token 
& committed token \\
\hline
\textbf{Output} 
& uncommitted 
& committed 
& committed \\
\hline
\textbf{Description} 
& low confidence 
& high confidence 
& update KV cache \\
\bottomrule
\end{tabular}
}
\end{table}
Each token within a block can be in one of three states, as summarized in \Cref{tab:states}. Both \emph{masked} and \emph{decoding} tokens are initialized with a fixed mask token. During each step, the model predicts token values and assigns confidence scores. Tokens whose confidence exceeds a threshold are committed (Decoding $\rightarrow$ Decoded), while others remain masked.

Although decoding tokens produce tentative outputs, their KV states are computed from masked inputs and are therefore inconsistent with the final committed tokens. Once a token is committed, it must be recomputed using the committed token to produce correct KV states. As a result, \emph{decoded} tokens correspond to KV states generated from committed tokens, while \emph{decoding} tokens only provide provisional states.

Increasing the block size expands the decoding window, allowing the model to evaluate multiple token positions simultaneously. Within this window, some positions may reach the threshold earlier and be committed ahead of others (e.g., tokens 1, 2, 5, and 6 in \Cref{fig:chunk}(a)). This is the fundamental reason why diffusion decoding improves throughput.

However, this parallelism introduces complex intra-block dependencies and rigid execution boundaries. Our goal is to break this rigid execution model by enabling finer-grained execution within each block.

\subsection{Eliminating Prefix Dependency via Caching}
A major source of dependency in block-wise diffusion decoding comes from repeatedly updating the hidden states of the decoded token. This creates strong prefix dependencies across decoding steps.
Prior work~\cite{ma2025dkv,wu2025fastv2,liu2025dllm} has shown that most decoded tokens and their KV states quickly stabilize after being committed and rarely change in later steps within the same block. As a result, recomputing their KV cache is largely unnecessary. Based on this, prefix caching reuses the KV cache of decoded tokens and skips their recomputation in subsequent steps~\cite{ma2025dkv,liu2025wedlm}. In \Cref{fig:chunk}(b), prefix decoded tokens can be excluded from computation.

However, this reduction in computation does not directly translate into performance gains without system support. Under low load, reducing computation does not improve throughput due to underutilized GPU resources. Under high load, irregular decoded token numbers across requests introduce additional overhead, which can offset the benefits.

Despite these limitations, prefix caching provides key abstraction: it removes prefix dependencies by decoupling decoded tokens from subsequent computation. Conceptually, this extends inter-block caching into the intra-block phase and enables excluding prefix tokens from execution.

\subsection{Reducing Suffix Dependency via Chunking}
While prefix dependency can be eliminated via caching, suffix tokens introduce another challenge. Fortunately, suffix tokens can be reduced without affecting correctness.
Prior work~\cite{wu2025fastv2,chen2025dpad,hu2025accelerating,wu2025fast,wang2025diffusion,li2025prophet} has shown that suffix tokens in diffusion decoding do not carry committed semantic content, and their computation can be cached~\cite{wu2025fast,hu2025accelerating, wu2025fastv2}, or partially skipped~\cite{chen2025dpad,li2025prophet,wang2025diffusion}. The primary role of suffix tokens is to provide a larger decoding window, allowing DLLM to evaluate multiple token positions simultaneously and increase the likelihood of \emph{``early commitment''}. For example, tokens such as positions 1, 2, 5, and 6 in \Cref{fig:chunk}(a) can be identified and committed earlier due to this expanded context.

Based on this observation, we reduce the effective decoding window into smaller execution units, which we refer to as 
\textbf{\emph{chunks}}. As shown in \Cref{fig:chunk}(c), chunked decoding partitions a block into smaller units, enabling flexible scheduling and fine-grained control over decoding granularity.

\subsection{Streaming Chunked Decoding}
Naively executing chunks in isolation can disrupt decoding order, increasing the number of decoding steps and degrading throughput. This explains why existing approaches do not directly adopt fine-grained chunking.
To address this, we introduce \emph{streaming chunked decoding}, shown in \Cref{fig:chunk}(d). The key idea is to dynamically reorganize chunks at each step to approximate the original decoding order.

Our design is based on three components:
\begin{itemize}[noitemsep, topsep=2pt, leftmargin=*]
    \item \textbf{Fine-grained caching}, which enables reuse of the KV cache at the chunk level;
    \item \textbf{Dynamic chunk sizing}, which supports multiple granularities and enables seamless switching between them;
    \item \textbf{Step-wise reorganization}, which dynamically adjusts chunk positions based on current decoded tokens and reconstructs chunk inputs at each step.
\end{itemize}

These mechanisms provide two key benefits. First, streaming execution converts otherwise unused prefix capacity into useful computation over suffix tokens. By shifting computation from prefix regions to undecoded regions, it maximizes the efficiency of each chunk.
Second, by expanding the effective decoding window and dynamically adjusting execution order, streaming chunked decoding closely approximates the original block-wise decoding schedule. Compared to naive chunking (\Cref{fig:chunk}(c)), this significantly reduces stalls at chunk boundaries and avoids unnecessary increases in decoding steps.
For example, in \Cref{fig:chunk}(d), the execution order of tokens closely matches that of block-wise decoding in \Cref{fig:chunk}(a), with only minor deviations (e.g., tokens 12 and 28). In contrast, naive chunking leads to substantial reordering, resulting in more decoding steps and lower efficiency.

This design departs from conventional diffusion decoding, which typically increases block size to expand the suffix region and improve the probability of early commitment. In contrast, chunked decoding reduces execution granularity and trades excessive parallelism for improved system efficiency.
Importantly, combining prefix caching and chunked decoding preserves model correctness. While reordering may introduce minor variations, our evaluation shows that accuracy remains stable and can even improve in some cases.
 We integrate streaming chunked decoding into existing DLLM serving pipelines and implement customized attention kernels to efficiently support dynamic chunking and KV reuse. 

Overall, chunked decoding integrates two key insights, prefix caching and suffix reduction, to enable dynamic scheduling. This not only improves system efficiency but also reshapes the execution paradigm of diffusion LLMs, potentially informing future algorithm design.

\section{Saturation-aware Elastic Scheduling}\label{sec:scheduling}

Chunked decoding enables fine-grained control over execution granularity, but selecting the optimal chunk size at runtime remains challenging. We address this by jointly modeling system and algorithm behavior to determine the optimal granularity at each step.

\subsection{System-Algorithm Co-modeling}
\begin{figure}[t]
    \centering
    \includegraphics[width=\linewidth]{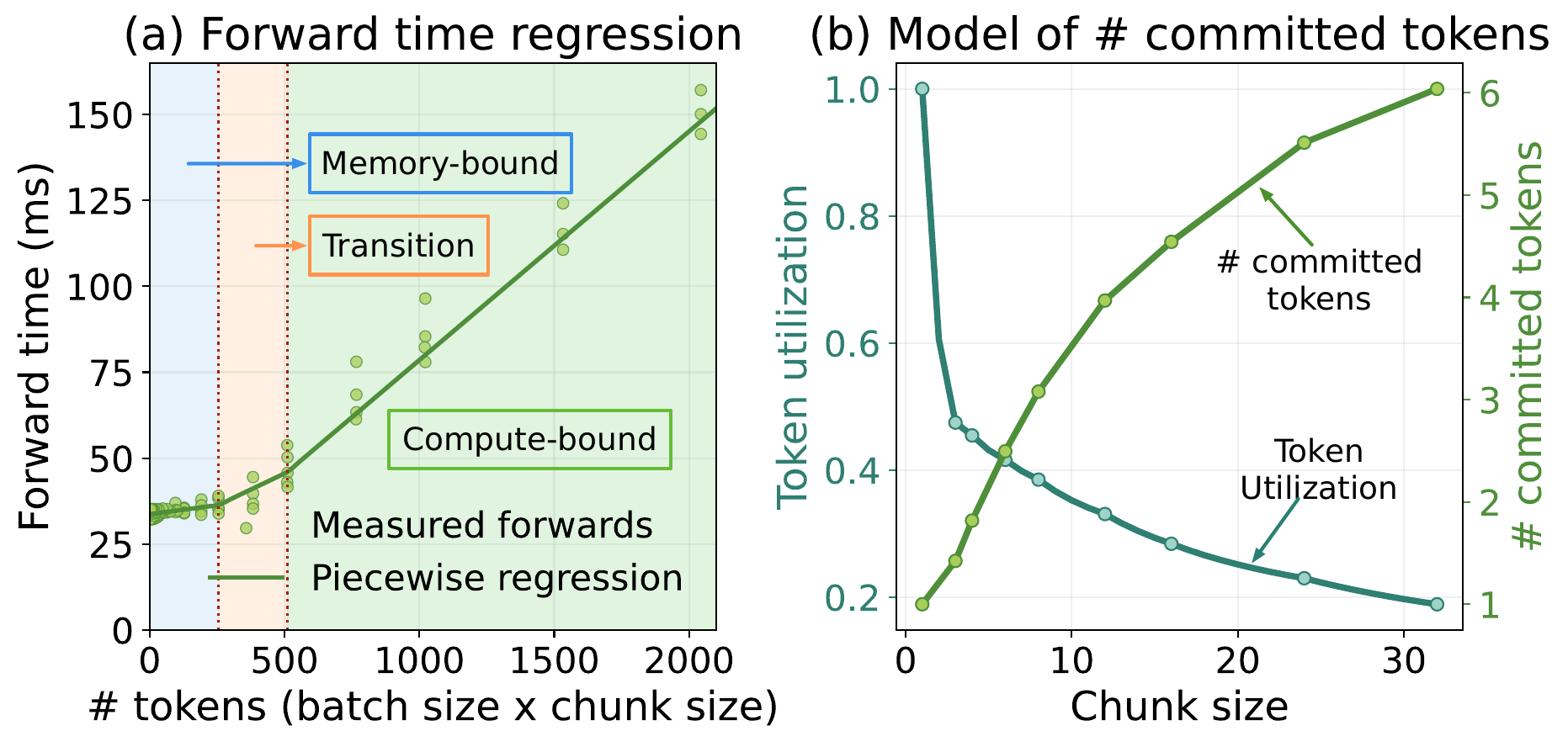}
    \vspace{-5mm}
    \caption{
    GPU latency and the committed token modeling.
    }
    \vspace{-5mm}
    \Description{}
    \label{fig:elastic}
\end{figure}

As illustrated in \Cref{fig:elastic}(a), for system efficiency, we build an offline estimator by profiling the target GPU and model across combinations of chunk size and batch size. At runtime, given the current batch size, we evaluate candidate chunk sizes and estimate their latency. As shown in \Cref{fig:elastic}(b), for algorithm efficiency, we estimate token utilization online for each chunk size, predicting the number of committed tokens of the current decoding stage.

Combining these two factors, we select the optimal chunk size $c^*$ at each step by solving:
\[
c^* = \arg\max_{c \in \mathcal{C}} \frac{N_{\text{commit}}(c)\times b}{T_{\text{latency}}(c, b)},
\]
where $\mathcal{C}$ is the set of candidate chunk sizes, $N_{\text{commit}}(c)$ is the estimated number of committed tokens for chunk size $c$, and $T_{\text{latency}}(c, b)$ is the estimated latency given chunk size $c$ and the current batch size $b$ obtained from continuous batching. We next describe the modeling of each component in detail.

\subsection{Modeling System Efficiency}

We model the decoding latency $T_{\text{latency}}(c, b)$ based on the dominant GPU workload. In practice, latency is primarily determined by the fully connected (FC) layers, whose computation scales with the total number of tokens processed per step, i.e., $b \cdot c$.

As illustrated in \Cref{fig:elastic}(a), the FC workload exhibits three regimes as $bc$ increases: a \emph{memory-bound} region at small $bc$, a \emph{transition} region, and a \emph{compute-bound} region at large $bc$. This behavior arises from the changing arithmetic intensity of FC layers: at small $bc$, computation is limited by memory bandwidth, while at larger $bc$, GPU compute becomes the bottleneck.

Although decoding attention also contributes to latency, it remains memory-bound and is not the dominant factor in our setting. We therefore approximate decoding latency using a piecewise affine model over $bc$:
\[
T_{\mathrm{latency}} \approx \beta^{(k)}_1 \, bc + \beta^{(k)}_0, \quad k \in \{1,2,3\},
\]
where each regime $k$ corresponds to memory-bound, transition, and compute-bound regions, respectively.

The coefficients $(\beta^{(k)}_1, \beta^{(k)}_0)$ are obtained via offline profiling. In the memory-bound regime, the slope is small and latency is dominated by fixed overhead, indicating low GPU utilization. As $bc$ increases, the slope grows in the transition region, and eventually becomes linear in the compute-bound regime, where latency scales proportionally with workload.

This piecewise model captures the non-ideal transition between memory- and compute-bound execution and provides a lightweight yet accurate estimator of GPU latency for guiding chunk-size selection in \proteus.

\subsection{Modeling Algorithm Efficiency}

In contrast to GPU latency, token utilization depends on model behavior and input data, and is difficult to predict offline. We define token utilization (TU) as
\[
\text{TU} = \frac{\text{\# committed tokens}}{\text{chunk size}}.
\]

As illustrated in \Cref{fig:elastic}(b), the number of committed tokens increases with chunk size but exhibits diminishing returns, leading to decreasing token utilization as chunk size grows. To estimate TU at runtime, we adopt an online approach. During early decoding steps, we observe the number of committed tokens under the largest chunk size (i.e., the standard block size), and construct an empirical mapping from chunk size to committed tokens, denoted as $N_{\text{commit}}(c)$. This estimate is continuously updated as decoding progresses, enabling the system to adapt to different inputs and workload characteristics.

Although the exact number of committed tokens is highly input-dependent and may deviate from our estimate in practice, the overall trend remains stable. Combined with our accurate system-level latency model, this approximation is sufficient to guide scheduling decisions, allowing \proteus to achieve strong performance in online serving scenarios.

\subsection{Saturation-aware Scheduling}

\begin{figure}[t]
    \centering
    \includegraphics[width=\linewidth]{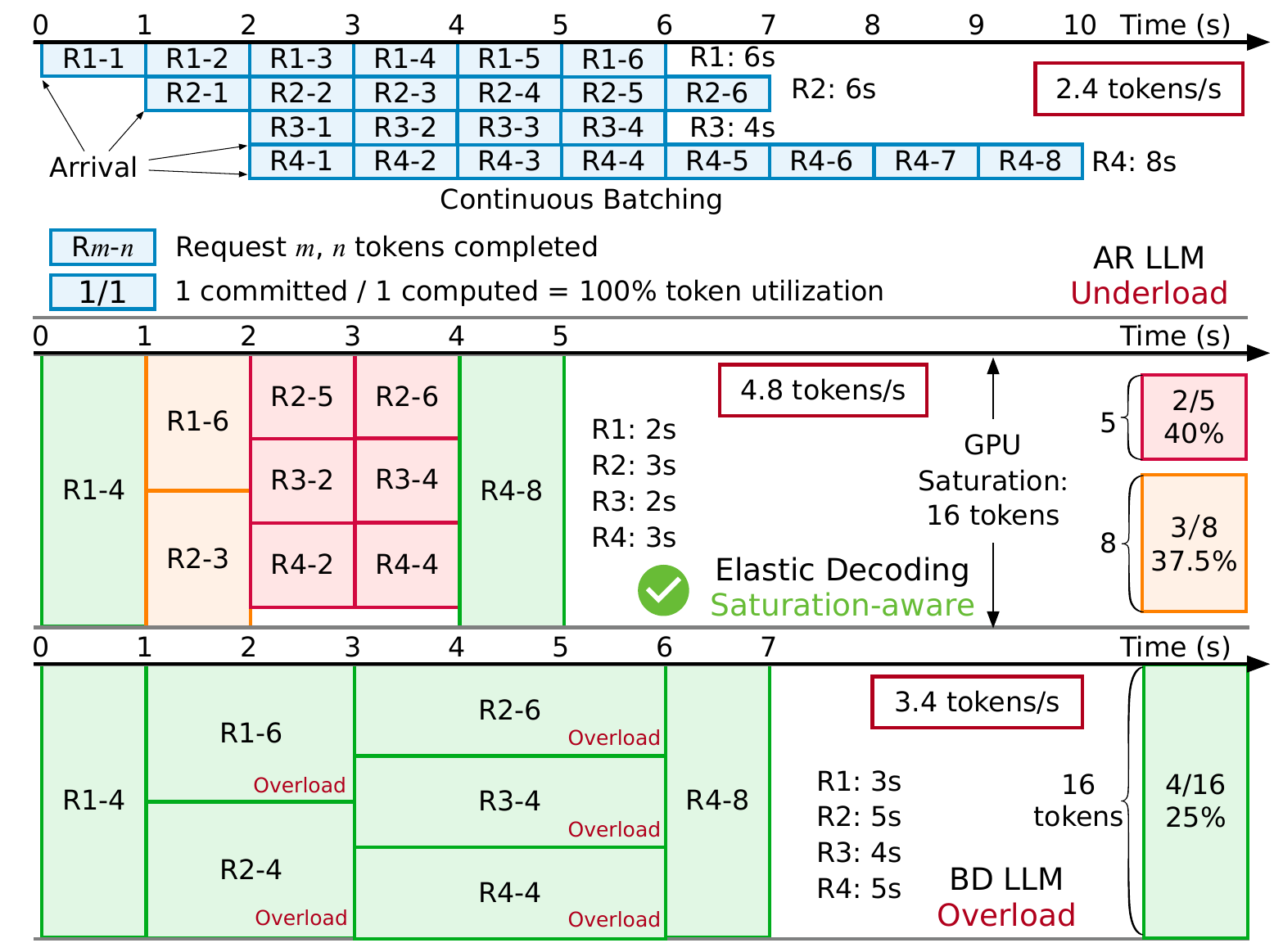}
    \vspace{-5mm}
\caption{
Scheduling comparison under a GPU capacity of 16 tokens (1 step = 1s under underload). 
Top: AR underutilizes the GPU. 
Middle: elastic decoding operates near saturation. 
Bottom: BD overloads the GPU. 
Elastic scheduling balances throughput and latency across workloads.
}
    \Description{}
    \label{fig:elastic_scheduling}
    \vspace{-10pt}
\end{figure}

We compare the execution behavior of three decoding strategies in \Cref{fig:elastic_scheduling}: autoregressive (AR) decoding, block diffusion (BD), and our elastic decoding.

\textbf{\textit{AR: Underutilization.}}
AR decoding struggles to reach GPU saturation. Under low-load conditions, each step exposes limited parallelism, leading to low throughput. Meanwhile, latency grows linearly with the number of decoded tokens, as each token must be generated sequentially. As shown in the top panel, AR achieves high token utilization but fails to efficiently utilize GPU resources.

\textbf{\textit{BD: Overload.}}
Block diffusion exhibits the opposite behavior. Without adaptive control, BD tends to overload the GPU, especially when batching is applied. This leads to two key issues.
First, batching under overload can increase latency. For example, the execution of blocks such as R1-6 and R2-4 becomes overloaded under contention, effectively doubling their execution time. As a result, R1 completes later than necessary. In contrast, if R1-6 were scheduled without batching, it could complete at 2s instead of 3s, while R2-4 would still complete in 3s. This shows that batching under overload can unnecessarily delay earlier requests without improving overall completion time. This illustrates that BD is inherently sensitive to batching and does not naturally align with standard serving schedulers.
Second, overload reduces overall efficiency. Larger block sizes introduce lower token utilization, and under saturation, this redundant computation becomes pure overhead. As a result, BD may suffer from reduced throughput despite higher nominal parallelism. These limitations help explain why diffusion-based decoding has not been widely adopted in production systems.

\textbf{\textit{Elastic Decoding: Saturation-aware Scheduling.}}
In contrast, elastic decoding dynamically adjusts execution to operate near the GPU saturation point. By jointly adapting chunk size and batching behavior, it avoids both underutilization and overload.
As shown in the middle panel, elastic scheduling maintains high hardware utilization while preserving efficient execution order. It achieves both lower latency and higher throughput compared to AR and BD. This is enabled by combining chunked decoding with runtime scheduling, allowing the system to balance algorithmic efficiency (token utilization) and system efficiency (GPU utilization).

\textit{This analysis highlights that efficient DLLMs serving requires co-design between algorithm and system. Elastic decoding provides a practical mechanism to bridge this gap, enabling diffusion models to achieve stable and efficient performance across varying workloads. This not only improves serving efficiency but also opens new directions for designing algorithms that are better aligned with system constraints.}

\section{Implementation}

\textbf{System Design and Kernel Support.}
We implement \proteus on top of LMDeploy~\cite{2023lmdeploy}, an open-source, production-grade LLM serving system with strong support for diffusion LLM inference. Building upon LMDeploy’s diffusion decoding runtime, we extend the scheduling and execution stack to support in-block caching and streaming chunked decoding, which together enable flexible decoding granularity.
To efficiently support chunked execution, we further implement a Triton-based paged-attention kernel. The kernel maps per-chunk KV states into the paged KV cache and enables attention over variable-length query tokens, allowing seamless support for arbitrary chunk sizes rather than being restricted to single-token or fixed block-size execution.

\textbf{Runtime Profiling and Scheduling.}
\proteus employs a lightweight warmup phase to collect GPU profiling data and characterize the decoding behavior of the target diffusion model under the incoming workload. Based on these profiling results, the runtime elastic scheduler dynamically adjusts chunking policies to track the optimal operating point under changing load.
Overall, \proteus comprises approximately 8K lines of code and supports widely used diffusion LLM families, including LLaDA2.0~\cite{bie2025llada2} and SDAR~\cite{cheng2025sdar}.

\begin{table}[t]
\centering
\caption{Datasets characteristics.}
\vspace{-2mm}
\label{tab:length_unmask}
\resizebox{\linewidth}{!}{
\begin{tabular}{lcccc}
\toprule
\multirow{3}{*}{\textbf{Dataset}} & \textbf{Input tokens} & \textbf{Output tokens} &  
\multicolumn{2}{c}{\textbf{BD32 token/step}} \\
 & \multirow{2}{*}{mean(std)} & \multirow{2}{*}{mean(std)} & \multicolumn{2}{c}{mean(std)} \\
 &  &  & SDAR-8B & LLaDA2.0-16B \\
\midrule
\small{ShareGPT}      & 213 (508)  & 321 (214)  & 5.29 (9.44)   & 2.51 (4.19) \\
\small{LMSYS-Chat} & 89 (133)   & 183 (163)  & 4.81 (8.80)   & 2.52 (4.84) \\
\small{LongBench}     & 4015 (2057)& 116 (138)  & 6.06 (10.74)  & 1.63 (1.90) \\
\midrule
\small{GSM8K}         & 89 (22)    & 175 (67)   & 3.20 (5.68)   & 2.61 (4.07) \\
\small{HumanEval}     & 172 (65)   & 103 (62)   & 3.75 (5.96)   & 6.01 (8.51) \\
\small{MBPP}          & 155 (77)   & 49 (28)    & 1.96 (3.33)   & 3.34 (4.81) \\
\small{IFEval}        & 58 (24)    & 281 (264)  & 1.88 (3.90)   & 1.28 (1.74) \\
\bottomrule
\end{tabular}
}
\vspace{-6mm}
\end{table}

\section{Evaluation}

\subsection{Experiment Setup}

\noindent\textit{\textbf{Hardware.}} We run all experiments on NVIDIA A100-SXM4-80GB GPUs interconnected via NVLink.

\noindent\textit{\textbf{Models.}} We evaluate \proteus on two representative diffusion LLM families, LLaDA 2.0~\cite{bie2025llada2} and SDAR~\cite{cheng2025sdar}, covering a range of model sizes. Our main results use SDAR-8B, a dense model, and LLaDA2.0-16B, a Mixture-of-Experts (MoE) model, both configured with a standard block size of 32. We use a confidence threshold of 0.9 for decoding, according to common practice~\cite{bie2025llada2,cheng2025sdar}. To provide a stronger point of comparison, we also include the corresponding autoregressive base models (AR) from which these diffusion models are derived, namely Qwen3-8B \cite{yang2025qwen3} for SDAR-8B and Ling 2.0-16B~\cite{tian2025Ling2} for LLaDA 2.0-16B.
All experiments use FP16 weights and activations, following standard deployment practices. For the main models, we use a single GPU since they fit within the memory budget of one device. To evaluate scalability, we further experiment with larger models within the same families, where we enable tensor parallelism~\cite{shoeybi2019megatron} to support multi-GPU execution.

\noindent\textit{\textbf{Workloads.}} Token utilization of diffusion LLMs is workload-dependent; we evaluate \proteus on a diverse collection of serving traces and benchmark datasets. For serving experiments, we use ShareGPT~\cite{sharegpt_unfiltered}, which captures realistic conversational workloads; LMSYS-Chat-1M \cite{zheng2023lmsys}, a large-scale chat dataset derived from real user interactions; and LongBench~\cite{bai2024longbench}, which emphasizes long-context generation and understanding. To evaluate model quality and enable direct comparison between autoregressive and diffusion decoding, we further include GSM8K~\cite{cobbe2021training} for mathematical reasoning, HumanEval~\cite{chen2021evaluating} for code generation, MBPP~\cite{austin2021program} for basic Python programming tasks, and IFEval~\cite{zhou2023instruction} for instruction-following. \Cref{tab:length_unmask} reports the input and output length statistics and number of decoded tokens per step of these datasets on diffusion LLM models. 
For online serving, we generate request arrival traces using a Poisson arrival process. We set service-level objectives (SLOs) based on application requirements, following common practice in prior LLM serving work~\cite{zhong2024distserve}. For interactive chat workloads (ShareGPT and LMSYS-Chat-1M), we adopt a relatively stringent 50ms time-per-output token (TPOT) SLO, as this is generally perceived as instantaneous for interactive applications~\cite{card2018psychology}. For long-context workloads (LongBench), we relax the SLO to 100ms TPOT, as these tasks are less sensitive to fine-grained token latency and prioritize throughput over immediacy.

\begin{figure}[t]
    \centering
    \includegraphics[width=\linewidth]{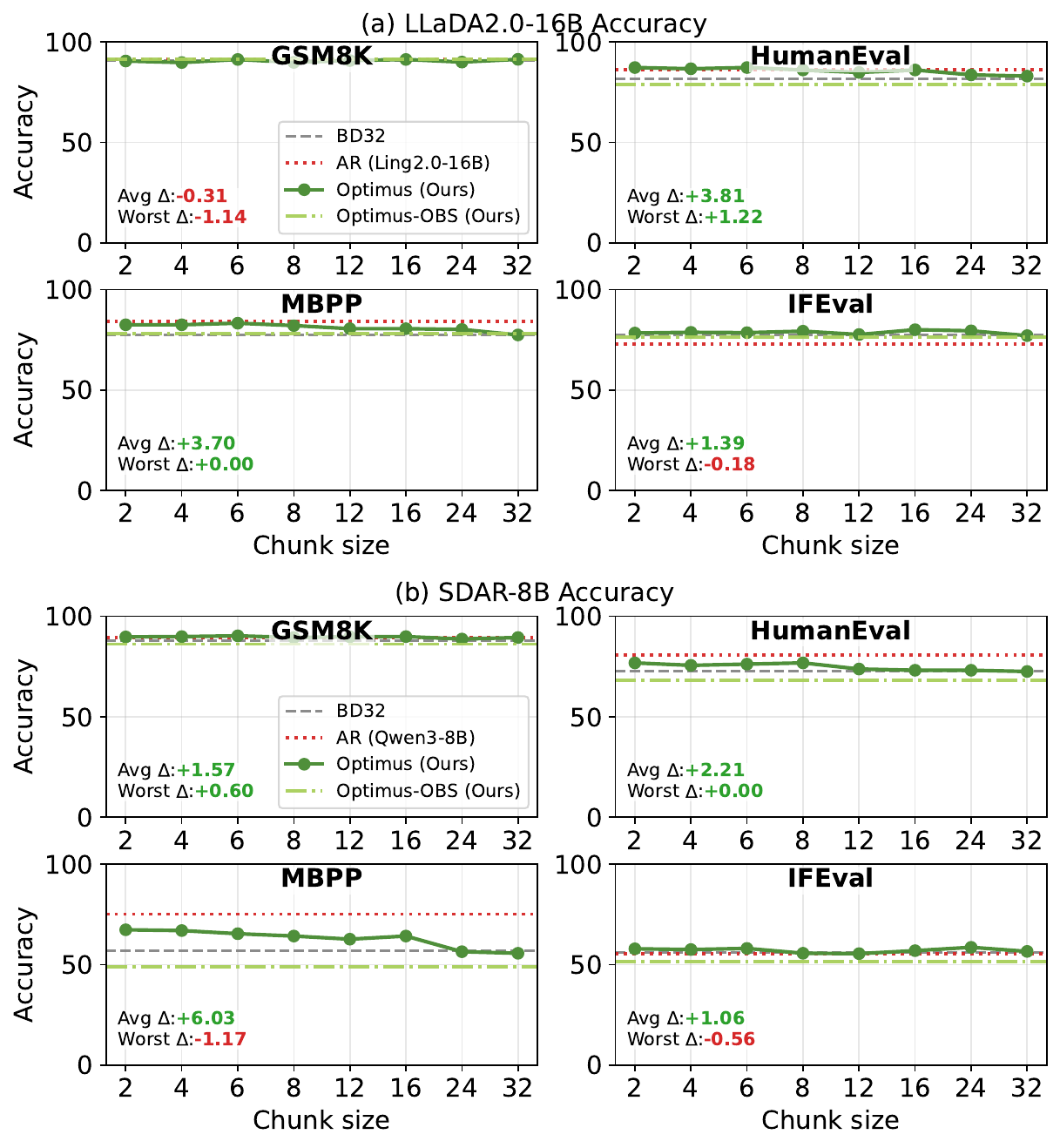}
    \vspace{-8mm}
    \caption{Model accuracy on common LLM benchmarks.
    }
    \Description{}
    \label{fig:accuracy}
    \vspace{-5mm}
\end{figure}

\noindent\textit{\textbf{Baselines.}}
We compare \proteus against two widely used serving frameworks with DLLM inference support. 

\textbf{SGLang}\footnote[1]{v0.5.9 Mar. 25, 2026 (commit ID 53c1d8e) }~\cite{zheng2024sglang} is a SOTA serving system designed for high-throughput LLM inference and is among the most widely adopted frameworks with official DLLM support. It extends standard LLM serving optimizations to the diffusion setting, including continuous batching and radix attention, enabling efficient batched execution. However, SGLang performs scheduling at a coarse granularity: batching decisions are made at the block level, and the batch is updated only after all requests finish decoding the current block. This design limits flexibility under dynamic workloads and can lead to suboptimal utilization when requests progress at different rates. SGLang adopts a first-come-first-served (FCFS) policy and prioritizes prefill over decoding.

\textbf{LMDeploy}\footnote[2]{v0.11.0 Dec. 12, 2025 (commit ID 32f1f0c)}~\cite{2023lmdeploy} is another high-performance serving framework for diffusion LLMs. Compared to SGLang, it employs a finer-grained scheduling strategy by updating batches at every decoding iteration, enabling more flexible continuous batching and improved resource utilization. LMDeploy also provides an optimized block-wise paged-attention kernel, making it a strong and efficient baseline for diffusion LLM inference. Similar to SGLang, it adopts FCFS scheduling with prefill prioritization.

To evaluate improvements over autoregressive decoding, we additionally include an \textbf{AR baseline} using LMDeploy. Since \proteus is implemented on top of LMDeploy, this comparison ensures a controlled setting where both systems share the same scheduling policy, runtime, and kernel implementations, isolating the impact of our proposed chunked decoding and elastic scheduling mechanisms.

\begin{figure}[t]
    \centering
    \includegraphics[width=\linewidth]{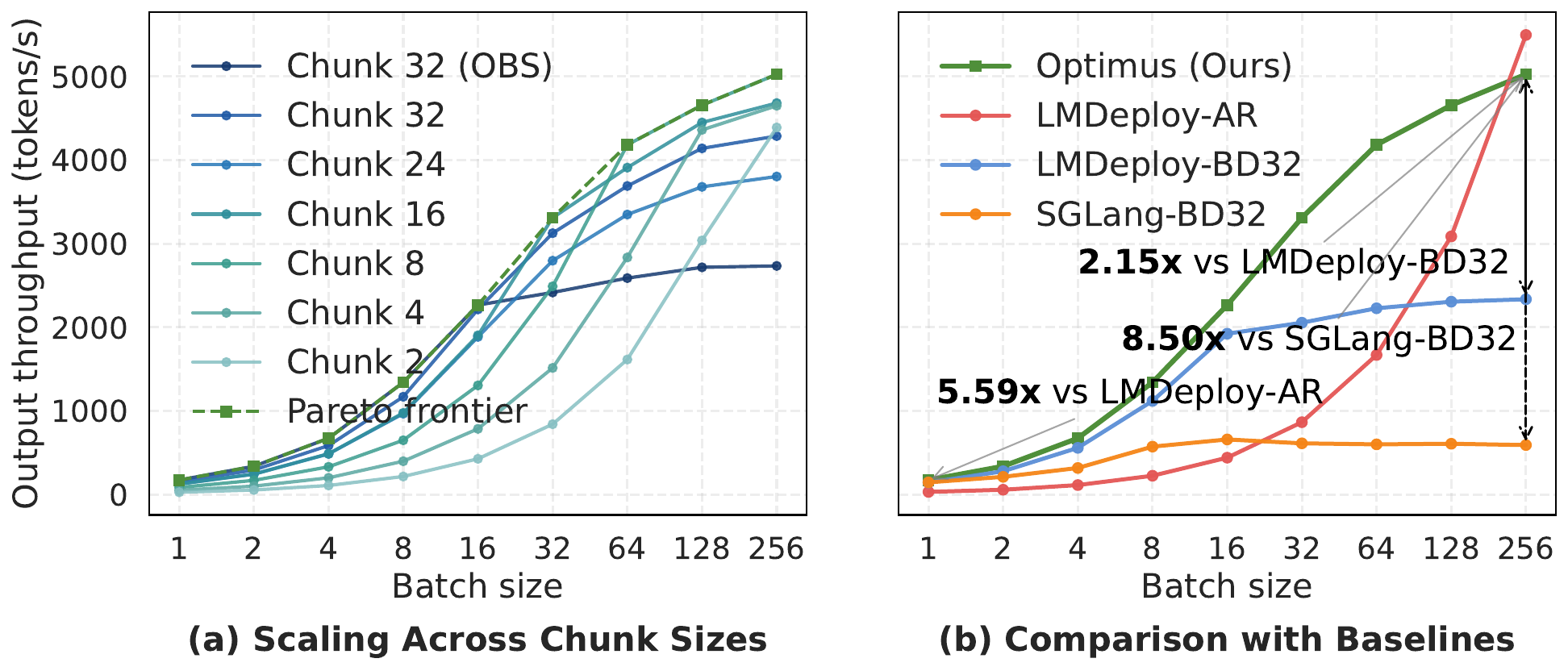}
    
    \vspace{-3mm}
    \caption{Throughput scaling with batch size. (a) Different chunk sizes exhibit a load-dependent trade-off and form a Pareto frontier; (b) \proteus adapts chunk size at runtime and outperforms AR and fixed-block baselines.
    }
    \Description{}
    \vspace{-10pt}
    \label{fig:throughput_fix}
\end{figure}

\begin{figure*}[!t]
    \centering
    \includegraphics[width=\linewidth]{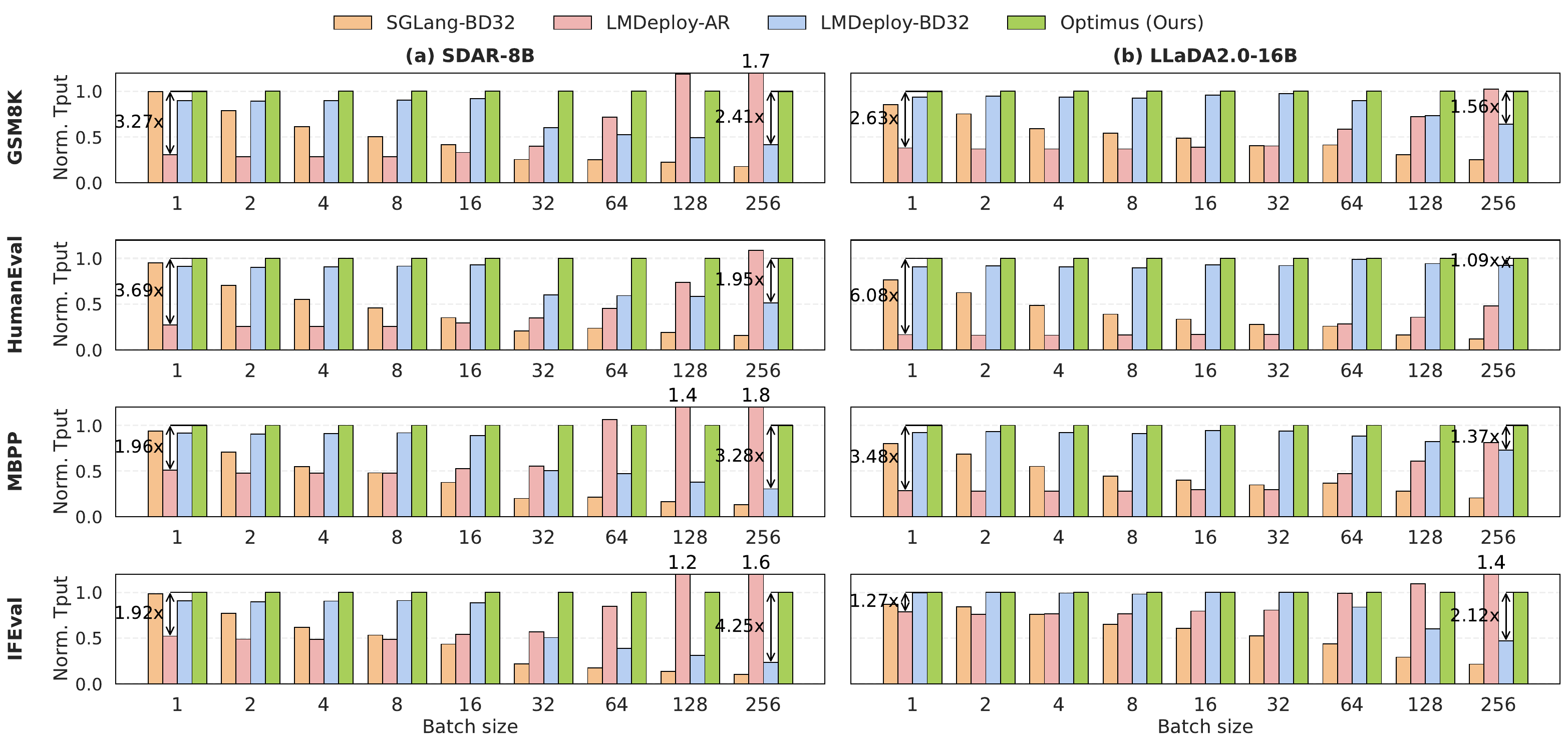}
    \vspace{-3mm}
    \caption{Throughput comparison across batch sizes.
    }
    \Description{}
    \vspace{-10pt}
    \label{fig:throughput_datasets}
\end{figure*}

\subsection{Model Accuracy}

We first evaluate whether \proteus affects model quality.
To preserve model accuracy, we adopt streaming only within a block, forbidding decoding tokens outside the current block, termed in-block streaming.
We measure accuracy across multiple benchmarks under varying chunk sizes, as shown in \Cref{fig:accuracy}. Overall, \proteus incurs minimal accuracy loss compared to standard diffusion decoding (BD32). The worst-case degradation across all chunk sizes is small, and in some cases \proteus even slightly outperforms BD32, indicating that modifying decoding granularity through chunked execution does not significantly impact model semantics.
Interestingly, we observe that the average accuracy across different chunk sizes often exceeds BD32 and approaches AR performance as the chunk size decreases. We attribute this behavior to the reduced decoding window in chunked decoding, which makes execution more AR-like.

We further evaluate a more aggressive out-block streaming (OBS) variant at chunk size 32, allowing streaming across blocks. While OBS can achieve higher throughput by allowing a larger decoding scope, particularly under low-load conditions, it introduces slightly larger accuracy degradation. This result highlights a controllable trade-off between efficiency and model quality, which \proteus can flexibly navigate this trade-off through its decoding configuration.
We attribute the accuracy degradation of OBS to the violation of block-wise dependencies assumed during training, as cross-block streaming alters the original decoding structure. In practice, we only enable OBS for large chunks (e.g., 32), while disabling it for smaller chunks where its impact on both performance and accuracy is limited. This also suggests potential directions for improving diffusion model design.

\subsection{Throughput Scaling with Batch Size}

We conduct a detailed analysis to understand how chunked decoding behaves under different load conditions, which manifest as varying batch sizes. Since our focus is on decoding efficiency, we report throughput measured in output tokens per second, excluding the prefill phase. 

To systematically study this behavior, we fix the batch size and sweep it from 1 to 256, while varying the chunk size from 2 to 32, including an additional configuration with a chunk size 32 using out-block streaming (OBS). We evaluate SDAR-8B on the ShareGPT workload. As shown in the \Cref{fig:throughput_fix} (left), no single chunk size is optimal across all batch sizes. Larger chunks (e.g., chunk 32 with OBS) achieve higher throughput at small batch sizes by exploiting idle GPU resources. 
As the batch size increases, the optimal chunk size gradually shifts to smaller values (e.g., chunk 16 and then chunk 8), reflecting the growing importance of token utilization. \proteus adapts to these changes and selects the best-performing chunk size for each batch size, achieving near-optimal throughput across the entire range.

We compare \proteus with AR and standard block diffusion baselines in \Cref{fig:throughput_fix} (right). \proteus outperforms all baselines in nearly all settings. It achieves a 5.59$\times$ speedup over AR at batch size 1, and 2.15$\times$ and 8.50$\times$ speedup over LMDeploy-BD32 and SGLang-BD32 at batch size 256.
AR slightly outperforms \proteus by 9.3\% at batch size 256 due to the property of diffusion decoding: tokens are predicted from masked positions, requiring an additional computation round to update the KV cache. As a result, each token is effectively computed at least twice, and the minimum feasible chunk size is 2 rather than 1. However, such high concurrency is uncommon in practical online serving, where \proteus consistently achieves superior performance.

We further evaluate throughput across a range of benchmark datasets 
and two DLLM models in \Cref{fig:throughput_datasets}. Despite differences in workload characteristics and token utilization patterns across datasets and models, \proteus consistently achieves higher throughput across all settings. These results demonstrate its ability to effectively balance GPU utilization and token efficiency across diverse workloads. Overall, \proteus achieves a geometric mean throughput improvement of 2.07$\times$ (up to 6.08$\times$) over LMDeploy-AR, 1.31$\times$ (up to 4.25$\times$) over LMDeploy-BD32, and 2.55$\times$ over SGLang-BD32 (up to 9.69$\times$) across all settings.

\subsection{End-to-End Serving Performance}

\begin{figure*}[t]
    \centering
    \includegraphics[width=\linewidth]{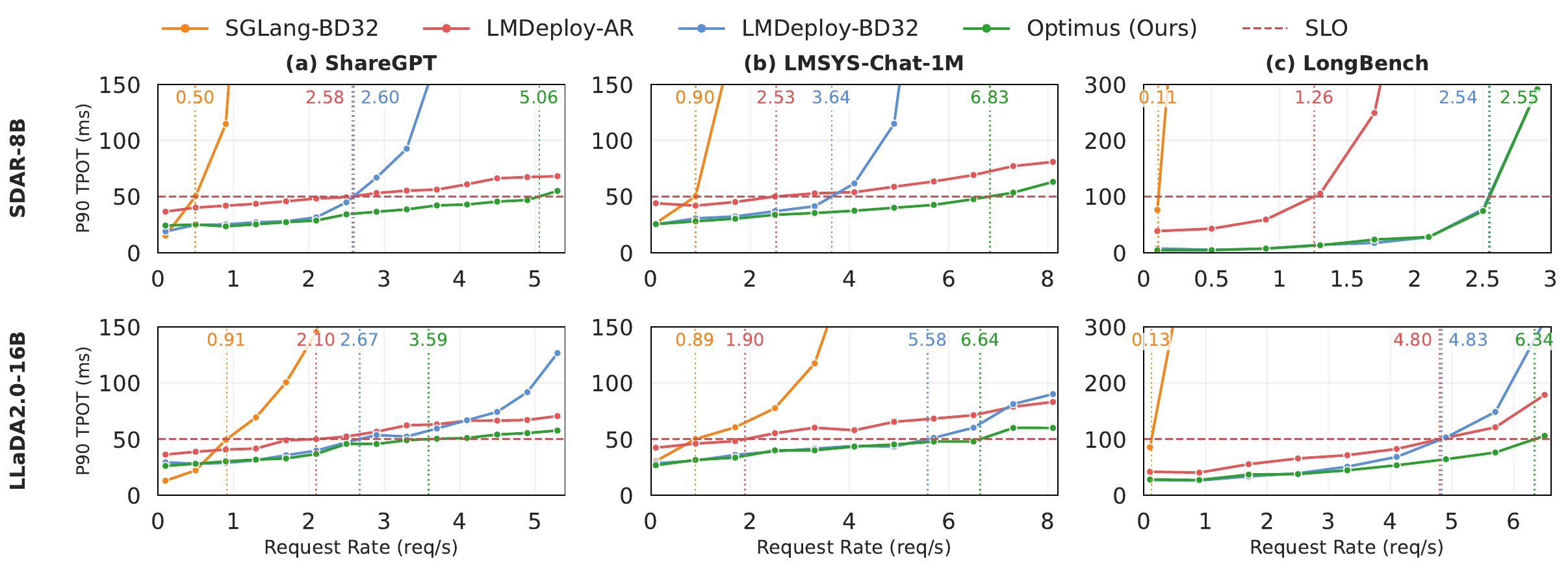}
    \caption{End-to-end online serving evaluation: P90 TPOT over different request rate.
    }
    \Description{}
    \vspace{-10pt}
    \label{fig:tpot_rate}
\end{figure*}

We next evaluate \proteus in an end-to-end online serving setting, focusing on decoding performance since the prefill behavior is largely identical across methods. Following prior work~\cite{zhong2024distserve,zhu2025nanoflow}, we measure the P90 TPOT under gradually increasing request rates in \Cref{fig:tpot_rate}. 

Overall, \proteus consistently achieves lower TPOT than the baselines across most workloads and load conditions. It also sustains substantially higher request rates under the same SLO. On SDAR-8B with ShareGPT, \proteus improves the SLO-compliant request-rate capacity by 10.2$\times$, 1.96$\times$, and 1.95$\times$ over SGLang-BD32, LMDeploy-AR, and LMDeploy-BD32, respectively. These show that \proteus translates its decoding-level gains in practical online serving.

Among various models and datasets, several common trends emerge. Compared with SGLang-BD32, \proteus performs similarly at extremely low request rates (e.g., 0.1 req/s), where the batch size is typically close to one. As the request rate increases, SGLang-BD32 becomes less efficient because its coarse-grained block-level scheduling leads to excessive redundant computation when requests within the same batch progress at different speeds. 
\proteus updates the batch at a finer granularity and dynamically adjusts chunk size. Compared with AR decoding, \proteus delivers lower TPOT across nearly the entire request-rate range, as real-world workloads rarely provide enough concurrency for AR decoding to fully utilize the GPU. \proteus, by contrast, exploits larger decoding granularity under low and moderate load to better utilize available GPU resources. Finally, compared with LMDeploy-BD32, \proteus shows similar TPOT under low request rates, where batch sizes remain small and large chunk sizes are still effective. As the rate increases, the fixed block size of 32 leads to GPU oversaturation, causing TPOT to rise rapidly. \proteus adapts by selecting smaller chunk sizes under higher load to maintain better token generation speed. \proteus improves end-to-end serving capacity by $2.1\times$ (up to $3.5\times$) over AR and $1.4\times$ (up to $2.0\times$) over BD32, and $11.8\times$ (up to $50.7\times$) over SGLang DLLM implementations.

\subsection{Runtime Scheduling Behavior Analysis}

\begin{figure}[t]
    \centering
    \includegraphics[width=\linewidth]{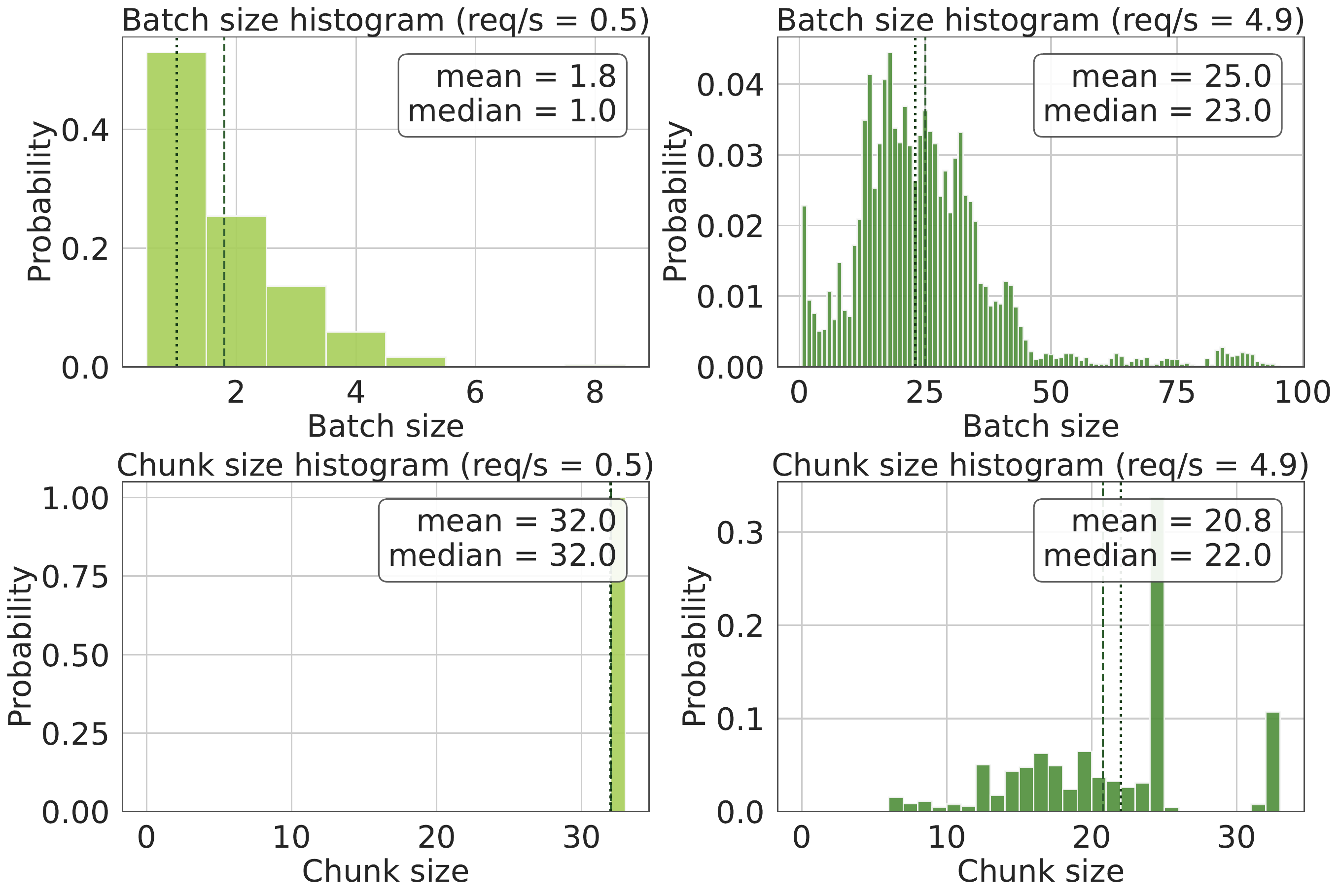}
    \caption{Batch size and chunk size distributions under low (0.5 req/s) and high (4.9 req/s) request rates.}
    \Description{}
    \label{fig:batch_distribution}
    \vspace{-15pt}
\end{figure}

To better understand the behavior of \proteus during online serving, we examine the runtime distributions of batch size and chunk size selected by the elastic scheduler. \Cref{fig:batch_distribution} presents SDAR-8B on ShareGPT under two representative request rates: low-load (0.5 req/s) and high-load (4.9 req/s).
Under low load, the batch size remains small throughout execution (mean 1.8, median 1). In this regime, \proteus almost always selects the largest chunk size 32. This is because the system is far from GPU saturation, and larger chunk sizes maximize GPU utilization and decoded tokens per step.

Under high load, the runtime behavior changes substantially. The batch size distribution shifts to much larger values (mean 25.0, median 23), occasionally approaching 100. Correspondingly, \proteus reduces the chunk size dynamically to avoid excessive redundant computation. In this setting, the selected chunk size averages 20.8 (median 22), while spanning a broad range and dropping to as low as 6 in some iterations. These results show that \proteus actively adapts its decoding granularity to the current system load, using large chunks to improve GPU utilization under low load and smaller chunks to preserve token efficiency under high load.

\subsection{Scalability Across Models and Tensor Parallelism}

\begin{figure}[t]
    \centering
    \includegraphics[width=\linewidth]{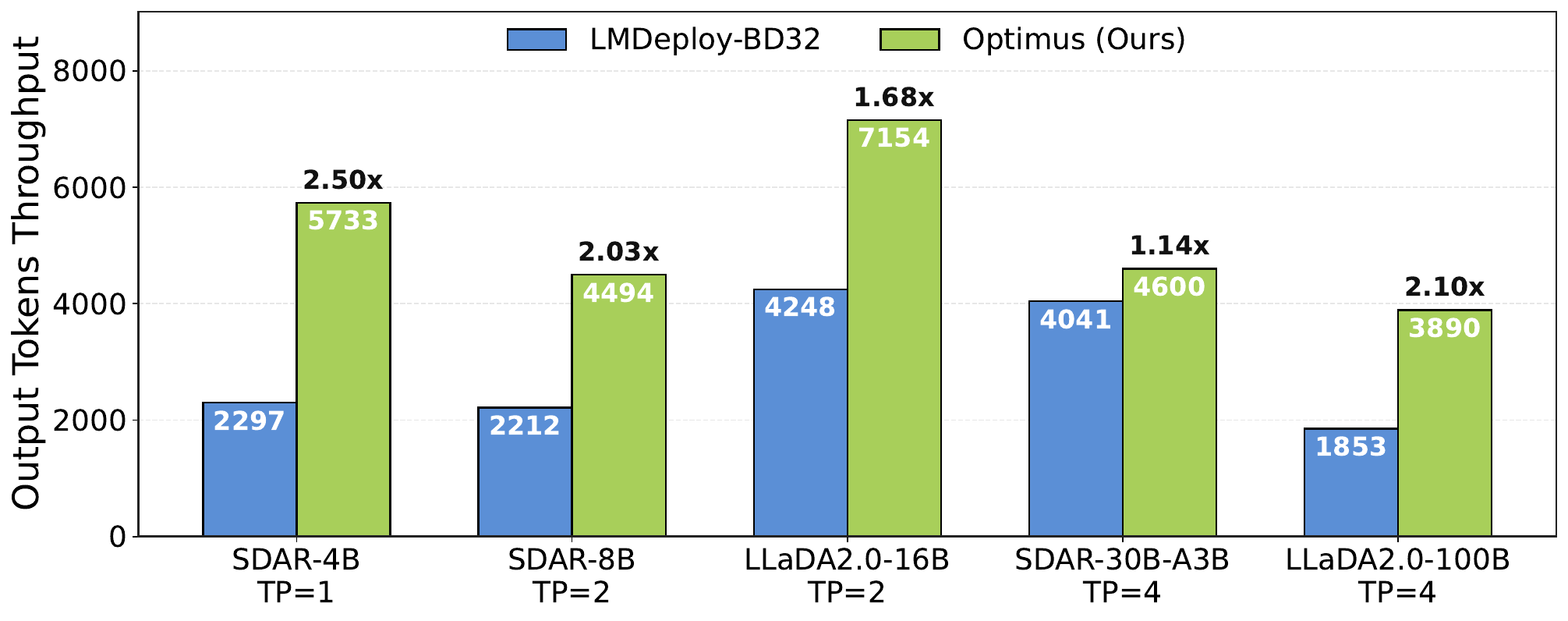}
    \caption{Throughput across model scales and tensor-parallel settings. 
    }
    \Description{}
    \label{fig:other_models}
    \vspace{-15pt}
\end{figure}

Finally, we evaluate the generality and scalability of \proteus across different model sizes and tensor-parallel configurations on GSM8K. As shown in \Cref{fig:other_models}, \proteus consistently outperforms the BD32 baseline in output-token throughput across models ranging from small variants to 100B parameters. The performance gains also persist under different tensor-parallel (TP) settings, as \proteus operates orthogonally to tensor parallelism. These results demonstrate that \proteus generalizes across model scales and deployment configurations, and can be incorporated into large-scale production serving systems while retaining its performance benefits.

\subsection{Ablation Study}

\begin{figure}[t]
    \centering
    \includegraphics[width=\linewidth]{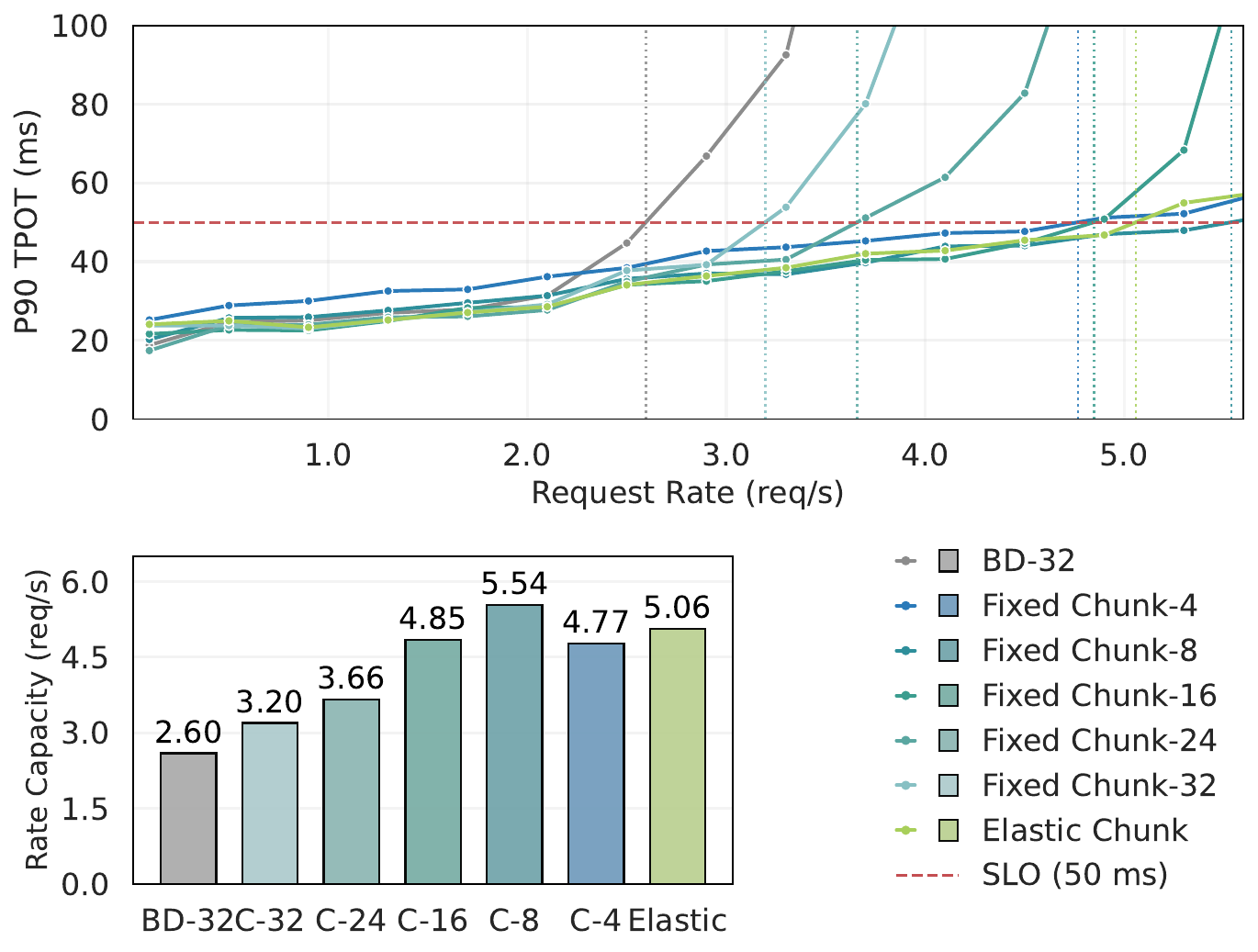}
    \caption{Ablation of chunked decoding and elastic scheduling.
    }
    \label{fig:ablation}
    \Description{}
    \vspace{-15pt}
\end{figure}

To understand the contribution of chunked decoding and elastic scheduling, we perform an ablation study by disabling these components individually. We first remove both mechanisms, which reduces \proteus to standard block diffusion decoding with block size 32 (BD32). We then enable chunked decoding along with fixed chunk sizes, without elastic scheduling, to isolate the effect of chunk granularity. Finally, we evaluate the full \proteus design, which combines chunked decoding with runtime elastic scheduling. We report P90 TPOT on ShareGPT using SDAR-8B in \Cref{fig:ablation}.

Chunked decoding alone already provides higher SLO-compliant request-rate capacity over BD32 in all fixed chunk sizes, and the best fixed configuration (Chunk-8) improves capacity from 2.60 to 5.54 req/s (2.13$\times$). This confirms that finer granularity reduces the redundant computation of large-block diffusion decoding. Adding elastic scheduling further improves robustness across request rates. As shown by the TPOT curves, \proteus maintains latency close to the best fixed-chunk configurations over a wide range of loads, indicating that the scheduler is effective at adapting chunk size online. The elastic policy reaches 5.06 req/s, outperforming BD32 and most fixed chunk sizes, and coming within 9.5\% of the best fixed configuration (Chunk-8). Given that the optimal chunk size depends on dynamic and difficult-to-predict decoding behavior, this result demonstrates that \proteus can effectively approximate the best static choice without requiring offline tuning for datasets and models.

\section{Related Works}
\textbf{\textit{LLM Inference Serving.}}
Prior work on LLM serving improves efficiency through scheduling, memory management, and system design within the autoregressive decoding paradigm. Orca~\cite{yu2022orca} introduces iteration-level scheduling for dynamic batching. vLLM~\cite{kwon2023vllm} improves throughput via PagedAttention and efficient KV-cache management. SGLang~\cite{zheng2024sglang} extends efficient serving with Radix attention for better KV reuse. Sarathi-Serve~\cite{agrawal2023chuckprefill} improves throughput--latency trade-offs through chunked prefill. DistServe~\cite{distserve} disaggregates prefill and decode to improve goodput under latency objectives. These systems primarily exploit \emph{inter-request} parallelism, and may still suffer from low GPU utilization under low-load conditions. \proteus is complementary by improving \emph{intra-request} parallelism via diffusion-style decoding.

\noindent\textbf{\textit{Multiple-Token Decoding and Adaptive Serving.}}
Recent work on multiple-token decoding accelerates generation by advancing multiple tokens per iteration. Speculative decoding~\cite{leviathan2022spec} drafts tokens and verifies them in parallel, with extensions such as EAGLE~\cite{li2024eagle} and EAGLE-3~\cite{li2025eagle3} improving proposal quality. Systems including TurboSpec \cite{liu2025turbospec}, AdaSpec \cite{hu2025adaspec}, and AdaServe~\cite{li2025adaserve} further adapt speculation policies online. However, these methods target autoregressive models and do not apply to DLLMs, which operate with fixed block structures and exhibit a distinct trade-off between GPU utilization and redundant computation. 
\section{Conclusion}

We present \proteus, a saturation-aware elastic decoding system for diffusion LLM serving. We show that fixed-granularity decoding is inherently suboptimal under dynamic workloads, leading to either GPU underutilization or inefficient over-saturation.
\proteus enables runtime adaptation of decoding granularity through chunked decoding and elastic scheduling, balancing GPU utilization and token efficiency to operate near the saturation frontier.
Experiments show that \proteus achieves up to $6.1\times$ throughput improvement over autoregressive decoding, $4.3\times$ over fixed-block diffusion, and improves serving capacity by up to $3.5\times$ under SLO constraints, while maintaining stable accuracy.


\bibliographystyle{ACM-Reference-Format}
\bibliography{ref}

\end{document}